# Network Slice Instantiation for 5G Micro-Operator Deployment Scenarios

Idris Badmus[1], Marja Matinmikko-Blue[1], Jaspreet Singh Walia[2], and Tarik Taleb[1,2]
[1]Centre for Wireless Communication, University of Oulu, Oulu, Finland
[2]Department of Communications and Networking, Aalto University, Espoo, Finland

*Abstract*—The concept of network slicing is considered as a key part in the development of 5G. Network slicing is the means to logically isolate network capabilities in order to make each slice responsible for specific network requirement. In the same light, the micro-operator concept has emerged for local deployment of 5G for vertical specific service delivery. Even though micro-operator networks are expected to be deployed using 5G, most research on network slicing has been directed towards the description on the traditional Mobile Network Operator's (MNO) networks with little emphasis on slicing in local 5G networks deployed by different stakeholders. In order to achieve slicing in a micro-operator network, it is of vital importance to understand the different deployment scenarios that can exist and how slicing can be realized for each of these deployments. In this paper, the micro-operator networks described include closed, open and mixed network, and for each of these network, different deployment scenarios are established. The paper further proposes approaches for the configuration of Network Slice Instances (NSIs) using the Network Slice Subnet Instances (NSSIs) and other Network Functions (NFs) in a micro-operator network while considering the different deployments. The results highlight the possible deployment scenarios that can be established in a micro-operator network and how network slicing can be efficiently realized for the various local deployments.

## I. Introduction

The idea of having a completely new locally deployed 5G network to complement the traditional Mobile Network Operator (MNO) networks for vertical specifics service delivery is the basis of the recently proposed micro-operator concept [1]. The aim of a micro-operator network is to provide enough network flexibility, privacy, and customization in the network to serve the local needs of verticals such as manufacturing and healthcare. In the past, some approaches by MNOs have been presented to better serve the verticals, such as multi-tenancy [2] and the use of base stations deployed within verticals' premises in need for better network coverage and capacity [3], but neither of these can provide the network flexibility, isolation, privacy and customization required by specific verticals.

The concept of micro-operators has been described using different terms, such as an industrial vertical network [4], a neutral host network [5], or a private network [6]. The micro-operator concept is expected to attract new stakeholders and new business opportunities in the telecommunication industry [7]. While many opportunities have been presented in terms of the overview and deployment scenarios of micro-operator networks [8], they will become a reality along with the next generation 5G technology where network slicing will be a crucial technology for their deployment and customization.

Network slicing is described by 3GPP as a key feature and a backbone in the coming generation network [9], [10]. It involves breaking down a network system from a "one size fits all" to a set of logically created networks with appropriate isolation and resource optimization. With the Software Defined Networking (SDN) and Network Function Virtualization (NFV) technologies, slicing in a network can be achieved because of the flexibility within these technologies [11]–[13]. Network slicing involves both core network and RAN slicing and adds network complexity. One way of implementing RAN slicing using a specific set of configuration description that parameterize and feature resources based on the radio protocol layers in a RAN node is described in [14]. Most of the network slicing research has considered implementation in MNO deployed 5G networks [15]–[17], with little or no concern has been on local 5G micro-operator networks. Although the work in [8] highlights the merits of slicing in the micro-operator network, no research on the realization of network slicing in different deployment scenarios has been done so far.

In order to achieve slicing in a network, it is important to understand how the network functionalities in the access and core networks, needed in forming the Network Slice Instance (NSIs) and the Networks Slice Subnet Instances (NSSIs) of slices are established [9], [15]. Also, to efficiently describe network slicing in a micro-operator network, it is important to establish different deployment scenarios and their subcases that can exist in the network. When these classifications are established, the micro-operator will be able to translate communication services requirements of the tenants to network slice requirements.

This paper presents foundational classification of different deployment scenarios that can exist in micro-operator networks and further proposes different types of NSI orchestrations for the considered deployment scenarios based on 3GPP's network slicing description [9]. The rest of this paper is organized as follows. Section II presents slice orchestration in a 5G network, while the different deployment alternatives of a micro-operator's network are described in Section III. Section IV focuses on network slicing for the different deployment scenarios. The concluding section consist of a general summary of the work done so far and possible future contributions.

This work is supported by Business Finland in uO5G and MOSSAF project, and Academy of Finland in 6Genesis Flagship (grant no. 318927).

## II. SLICE ORCHESTRATION IN A 5G NETWORK

According to 3GPP description of slicing in a network [9], the Communication Service Management Function (CSMF) is responsible for translating communication service requirement into network slice requirements. The Network Slice Management Function (NSMF) is then responsible for management and orchestration of the Network Slice Instance (NSI), while the Network Slice Subnet Management Function (NSSMF) is responsible for management and orchestration of the Network Slice Subnet Instance (NSSI). The NSMF manages and orchestrates the network slice while establishing communications with the NSSMF, and the CSMF. The NSI is an entity that describes the end-to-end creation of a network slice in an operator's network. In forming a network slice, different instances may occur which depends on the NSSI. The NSSI contains Network Functions (NFs) belonging to either the Access Network (AN) or Control Network (CN) which can be used to describe the type of NSI that will be created in the network.

A NSI is expected to contain two or more NSSIs [9]. For example, in the instantiation of a slice instance, the NSI may contain functionalities from both AN and CN. AN and CN are separate NSSI in the network. AN represents the NSSI1 for the RAN network functions while CN represent the NSSI2 for the core network functions. NSSI1 and NSSI2 will be used to describe the network slice instance. Different types of NSI configuration can be described based on the relationships between the NSSIs. Also, different modes of slice instantiation can be used by a tenant i.e. group of users, or services requiring slices [15]. The first is the **request mode** where a tenant requests a slice to be formed; in this way the operator just allocates all requested resources needed for the formation of the slice. Throughout the slice life cycle, all operations within the slice are managed by the tenant. In the second method known as the **pre-defined mode**, the slice creation is pre-defined by the operator for the tenant. In this method, the management of the slice during the lifecycle will be handled by the operator's network.

When an NF or NSSI is shared between two slice instances, a shared constituent is formed. Whether a slice will have a shared constituent or not depends on the type of service the slice is responsible for in the network. This leads to the basis of our approach in describing network slicing for different local micro-operator networks. Hence, three types of NSI configurations can be described based on the type of relationships that can be formed from the NSSI making up the slice instances. These three types are introduced next.

- **Type 1 NSI configuration**

Here, a slice instance is formed with AN and CN NSSI without a shared constituent. This type of NSI will be responsible for tenants who require strictly low latency, high security, and high reliability and do not want any network resource to be shared with other slices in the network to avoid impact from other services. Fig. 1 describes such a configuration.

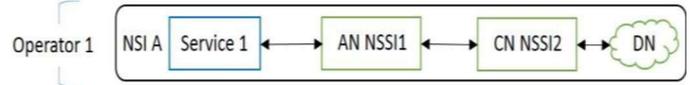

Fig 1. Type 1 NSI configuration.

- **Type 2 NSI configuration**

In this configuration, NSIs are formed with shared NSSI or NF constituent from other slices within the same network as described in Fig. 2. This type of slice instances can be created for tenants with less strict latency, security, and reliability requirement. However, sufficient amount of resource optimization is required with the shared constituents. Besides some of the shared NSSIs, different slices may also have their dedicated NSSIs.

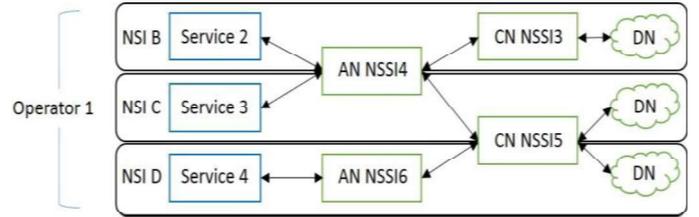

Fig 2. Type 2 NSI configuration.

- **Type 3 NSI configuration**

In Type 3 configuration, an NSI is formed with shared NSSI or NF constituent from slices in a different network. This configuration requires external network resources for services, such as wide area access, remote monitoring and roaming. Fig. 3 describes such a configuration.

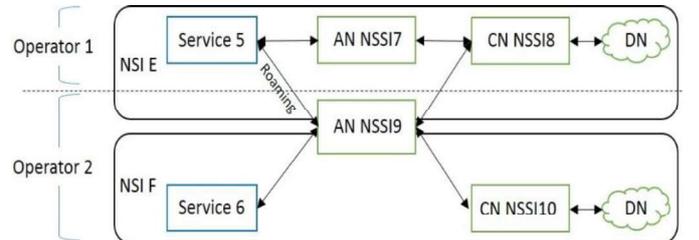

Fig 3. Type 3 NSI configuration.

Finally, it is important to know that depending on the vertical sector that is served by the 5G network and its involved use cases, one or more combination of these NSI types may be utilized by the operator. For example, in the industry vertical, low latency use cases could need Type 1 NSI orchestration, whereas broadband and less critical systems that require less stringent requirements could utilize shared NSSIs, i.e. Type 2 NSI orchestration, while logistic services could require Type 3 NSI configuration for wide area access. Table I summarizes the three NSI configuration types that can be considered for the different deployment scenarios in 5G operator networks.

TABLE I. NSI CONFIGURATION TYPES

| NSI Types | Configuration description |
|---|---|
| Type 1 | No shared NSSI constituent with other slice instances in the network, recommended for latency specific services by allocating a reserved slice. |
| Type 2 | There is shared constituent with slices in the same network, requires optimization of shared NSSI |
| Type 3 | There is shared constituent with slices from another network (MNO), required for specific service needed outside the micro-operator's network |

## III. 5G MICRO-OPERATOR DEPLOYMENT SCENARIOS

A micro-operator network [1], [5], [7], [8] is a local 5G network deployed for different purposes. It can be a closed, open or mixed network depending on the customers that it serves. For each of these networks, we consider different deployment scenarios that can be realized. **A closed network** [18] is a private network where the users and tenants require some authentication, for example, a private network covering an industrial area, a closed hospital network, or a private campus network without MNO subscribers. Here, the network resources can be allocated to individual tenants within the network and slice instances can be created for each tenant. The closed network in a micro-operator's network will generally have two deployment scenarios, namely: Deployment A and Deployment B. Deployment A is here defined as a closed micro-operator network serving one or more tenants with operations in a single location only. Deployment B denotes a closed network serving one or more tenants with operations in multiple locations. The communication for a tenant over multiple locations can be enabled by connecting different local networks or by sharing external subnet instances with another operator (e.g., MNO). The security and isolation requirements will be higher for multiple tenants served over the same network. Basically, the difference between Deployments A and B in the closed network is the location that the micro-operator is covering.

In the **open network,** [19] we consider two types of network deployments for a micro-operator, an MNO Open network and a Public Open network. The MNO open network is targeted at serving MNO subscribers within a locality which requires that the micro-operator has a service agreement with the MNO to be responsible for its customers within the given locality. For example, in a local area where building an MNO network is costly or complicated, the MNO can buy the service from a micro-operator that potentially serves multiple MNOs in that area. The micro-operator network can thus be responsible for subscribers from a single MNO (less security and isolation required) or be responsible for subscribers from multiple MNOs (more security and isolation required). Overall, the relationship is based on the type of agreement between the micro-operator and the MNO(s). The public open network on the other hand is a micro-operator network with lower level of authentication compared to the MNO open micro-operator network and users are not allocated specific network resources. It is similar to the general open Wi-Fi network whereby visitors can connect to the network without prior agreement with the micro-operator.

**The mixed network** is a micro-operator network covering both open and closed networks with predefined levels of authentication and isolation given between their different subscriber sets. This is done in such a way that the open part of the network covers MNO subscribers whom the micro-operator has an agreement with, i.e. the MNO open and the public open. The closed part of the network covers tenants with authentication and security. Hence, the mixed network will cover both closed and open networks, and all considered deployment scenarios. Additionally, based on the relationship with the MNO network, the mixed network deployment scenarios can be either Option A or Option B. Option A refers to the network scenario whereby the micro-operator is in need of network resources from an MNO network for wide area access, such as accessing content in faraway servers, remote monitoring, and roaming scenarios for multi-site operations.

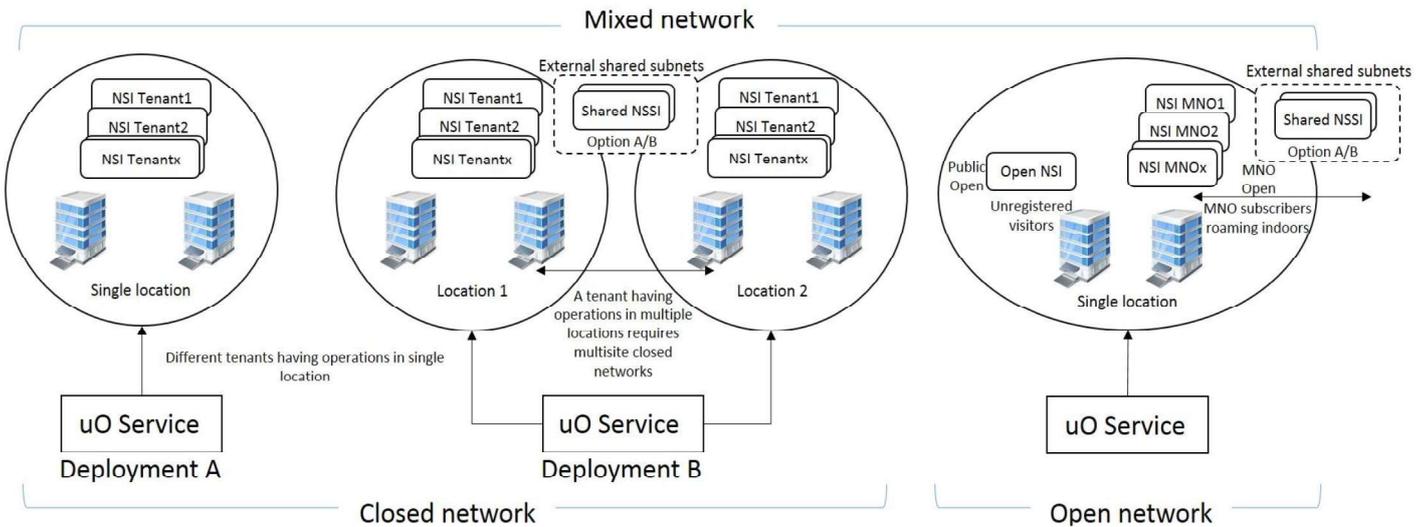

Fig 4. General overview of the of different micro-operator network deployments.

Option B is when the MNO needs access to the micro-operator's network to serve its subscribers within the micro-operator's network, similar to the MNO open case. However in this option, MNO wants to serve some of its own customers within the mixed micro-operator's network, or the MNO(s) is using the verticals broadband slice to extend their indoor coverage. Fig. 4 describes the general overview of the different micro-operator's network deployment scenarios.

## IV. NETWORK SLICING IN DIFFERENT MICRO-OPERATOR'S DEPLOYMENT SCENARIO

As explained in [8], network slicing in a micro-operator will be critical enabler for proper service delivery in the local network that can serve different customer sets. Within the local area of coverage for the micro-operator network, the advantage of lower latency due to the proximity of the network operator will be more efficiently supported with slicing in the network. Likewise, for a vertical, each slice requirement in the micro-operator network will be more service-specific, thereby bringing more efficiency and dynamics to the network. These advantages have prompted studies into how network slicing can be implemented for different deployments scenario in a local 5G micro-operator network. As highlighted in Section III, the micro-operator network will be considered to include closed, open and mixed networks, which calls for different network slice instantiations. The network slice instance (NSI) configuration for each of the deployment scenarios in these networks is based on Type 1, Type 2 and Type 3 NSI configuration as illustrated in Fig. 1, Fig. 2 and Fig. 3 respectively. The slicing description for different deployment scenario in the micro-operator's network are described next.

### A. The closed micro-operator network

**Deployment A** is described when one micro-operator network is closed and it is responsible for a single site covering one or more tenants within the same location. The operator will be responsible for converting only service requirements within the tenants into slice requirements and providing the necessary resources for each of the services. The NSI configuration for the Deployment A in a closed network can be either Type 1 when there is one tenant in the network or Type 2 when there are multiple tenants but not Type 3 because there is no communication with MNO network. Fig. 5 describes slices instances for Deployment A. Services such as remote surgery, or industrial automation [20] that requires low latency and high isolation can be carried out efficiently with this type of deployment.

In **Deployment B** for the closed network, the micro-operator covers tenants at different locations and hence the slice instances will be determined based on the services of individual tenants within the network at those locations. Generally, each tenant in the network can have different slice requirement and resources attributed to them and this will be based on their location and services.

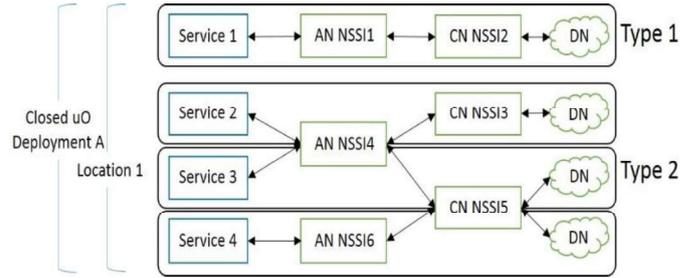

Fig 5. Deployment A: one closed network with both Type 1 and 2 NSI.

Unlike Deployment A, the Deployment B NSI configuration can be either Type 1, Type 2 or Type 3. Type 1 and Type 2 NSI configurations will cover the deployments when none of tenants within the close network will have any need to connect or share resources with the MNO network. This means the network is better secured since no connection will be made with any external operator.

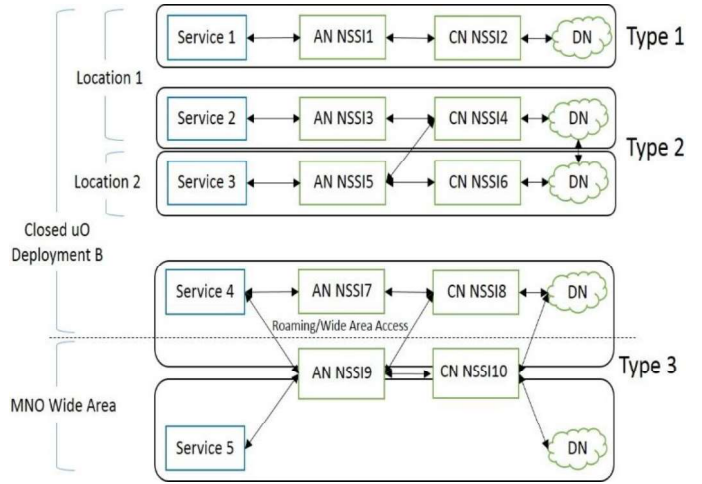

Fig 6. Deployment B: One closed network for multiple locations.

Deployment B of the closed network can be of a Type 3 NSI configuration when one or more tenants in the network have a need to connect to the MNO, thereby having shared constituent with the MNO network. This type of configuration happens when one or more tenants within the network requires services outside of the network, including services such as wide area access, roaming, inter-connection between two tenants at different location. Hence, better security is needed in the network. Fig. 6 describes such configuration for deployment B of the closed network.

### B. The open micro-operator network

The open network involves having a micro-operator network targeted at MNO subscribers (MNO open) and the general public (Public open). The NSI configuration type for the **MNO open** network depends on whether the micro-operator is responsible for subscribers of one MNO or Multiple

MNOs. If the micro-operator network is responsible for subscribers from only one MNO, the NSI Type 3 will be used to describe such configuration because the relationship is with a single MNO. Meanwhile, if the micro-operator is responsible for subscribers from more than one MNO, the Type 2 and Type 3 NSI configuration can be used and this can be seen in Fig. 7. Each slice from the micro-operator network is responsible for different MNO subscribers while maintaining a regular relationship with the MNO. Whereas in the same network, a separate slice is reserved for **Public open** subscribers which will have NSI Type 1 configuration because it does not share resources with the MNO subscriber's slice.

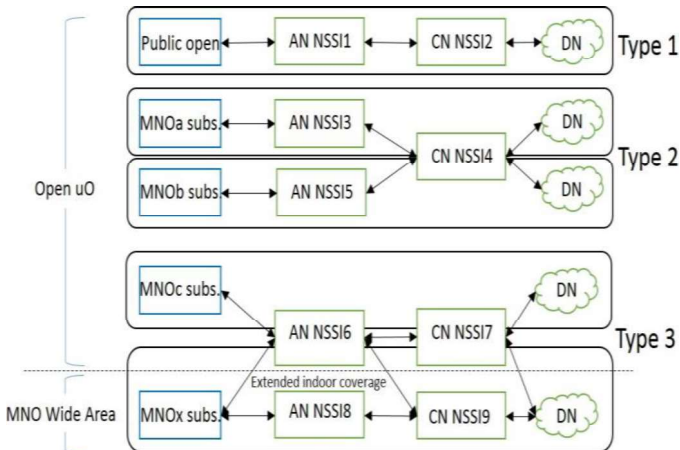

Fig 7. Open network slicing description.

*C. The mixed micro-operator network*

The mixed network will be the most comprehensive deployment of the micro-operators network. Therefore, it is of great importance to understand how network slice instantiation takes place within the network. In order to achieve proper slicing in the mixed network, the NSI configuration Type 1, Type 2 and Type 3 are applied with enough network security between the closed and the open parts of the network. Whilst the slices for the open part of the network aim at covering the MNO subscribers and the general public, the slices for the closed part serve private tenants with specific network requirements. Hence, since the mixed network covers all deployment scenarios of both open and closed network, there is a need for regular exchange with the MNO network to cover wide area services such as remote monitoring, wide area fleet maintenance for industrial verticals within the closed part of the network, and better indoor coverage of MNO network for its subscribers in the micro-operator's network. Meanwhile, even though the NSI configuration for the mixed network can be either Type 1, Type 2 or Type 3, the relationship between the MNO and the micro-operator will determine the deployment scenarios options we can have in the mixed network.

**Option A** as described in Fig. 8, takes place when micro-operator needs the MNO, thereby configuring the tenants' slices with the MNO resources for services that require wide area access. In this option the MNO can also extend its coverage indoors by utilizing micro-operator's resources. As shown in Fig. 8, some slice instances in the micro-operator network are made from both the MNO NSSI and the micro-operator NSSI, i.e. the micro-operator's NSSMF must have the capability to form NSI from micro-operator NSSI and MNO NSSI.

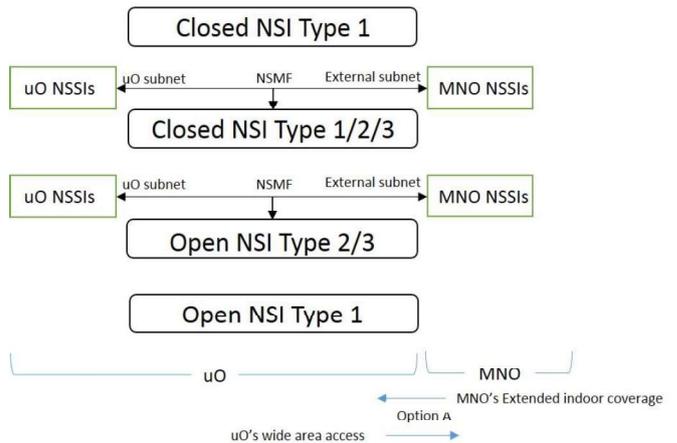

Fig 8. Mixed micro-operator network Option A, MNO NSSI

In contrast, **Option B** takes place when the MNO needs the micro-operator's network e.g. when MNO wants to expand its services or to provide better quality for its subscribers using the micro-operator's resources. This option is described in Fig. 9. It can be seen that the services are described with both the micro-operator NSI and the MNO NSI, i.e. the micro-operator's CSMF function must be able to manage the tenant's services consisting of both micro-operator and MNO slices.

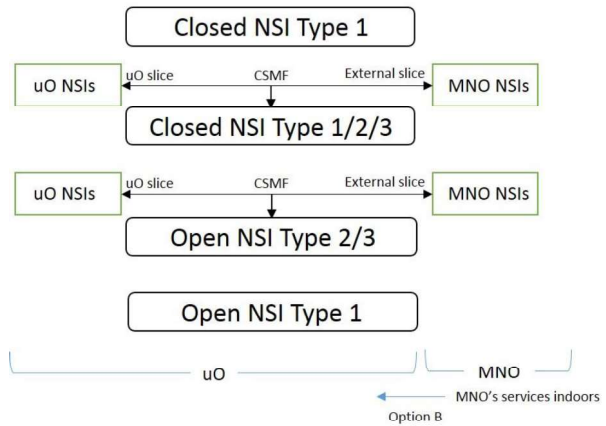

Fig 9. Mixed micro-operator network option B, MNO NSI.

It is expected that the mixed network will be the most comprehensive micro-operator's network because of its ability to cover both open and close scenarios within the same network. However, enough isolation and security are required to ensure that the closed part of the network remains closed without compromise. Table II summarizes the developed NSI configurations for the different deployment scenarios.

TABLE II. NSI CONFIGURATION FOR DIFFERENT MICRO-OPERATOR DEPLOYMENTS

| Micro-operator network | Deployment scenario | NSI configuration |
|---|---|---|
| Closed | Deployment A | Only Type 1 and Type 2 NSI configuration can happen because of no communication with MNO or other external network |
| Closed | Deployment B | NSI configuration can be Type 1 and Type 2 if no tenant in the network will connect to the external network and NSI configuration will be Type 3 if a tenant or two within the network will be connecting to the external network |
| Open | MNO Open | Type 3 NSI configuration if the micro-operator is responsible for subscribers from one MNO and Type 2 and Type 3 NSI configuration if the micro-operator is responsible subscribers from multiple MNO |
| Open | Public Open | A single NSI in the open network will be dedicated for the general public with Type 1 NSI orchestration. |
| Mixed | Option A | NSI configuration can be Type 1, Type 2 or Type 3 for the mixed network and the deployment configuration will have some of the slice instance made from MNO NSSI from both micro-operator NSSI |
| Mixed | Option B | NSI configuration can also be Type 1, Type 2 or Type 3. However, the services will be described with NSI from both micro-operator and MNO |

## V. CONCLUSION

This paper did not only lay down a solid foundation for the classification of different deployment scenarios of a local 5G micro operator network, but also described how network slicing can be efficiently implemented for each of the deployment scenarios. One of the main advantages of the micro-operator network is the amount of local customization it can offer to different verticals and their services, and network slicing in 5G is an efficient way to achieve this. This paper described different NSI configuration types that can exist in a micro operator network based on the 3GPP slice definition and paired the NSI configuration types to different deployment scenarios in the micro-operator network. From the classification made in this paper, a micro-operator can be either deployed as a closed, open or mixed network. The paper introduced different ways for NSI configuration of micro-operators leveraging network slicing. Network slicing in the closed micro-operator network may supports multiple tenants at different locations within a local micro-operator's network while maintaining a high level of privacy and isolation. Slicing in the open network may result in better network coverage for MNO networks. The mixed network enables both open and closed network with proper isolation between them. Based on the obtained results future research is needed to cover the practical implementation of selected deployment scenarios and the comparison with slicing in the MNO network.